\documentstyle[epsfig,axodraw]{aipproc}
\newcommand{\be}{\begin{equation}}
\newcommand{\ee}{\end{equation}}
\newcommand{\bea}{\begin{eqnarray}}
\newcommand{\eea}{\end{eqnarray}}
\newcommand{\al}{\alpha}
\newcommand{\gm}{\gamma}
\newcommand{\Gm}{\Gamma}

\newcommand{\eps}{\epsilon}
\newcommand{\ep}{\epsilon}

\newcommand{\ka}{\kappa}

\newcommand{\dd}{\mbox{d}}

\newcommand{\uQ}{\underline{Q}}
\newcommand{\uM}{\underline{M}}

\newcommand{\uq}{\underline{q}}

\newcommand{\um}{\underline{m}}

\newcommand{\nn}{\nonumber}

\begin{document}
\title{Asymptotic Expansions of Feynman Integrals
on the Mass Shell in Momenta and Masses}

\author{V.A. Smirnov$^*$}
\address{$^*$Nuclear Physics Institute of Moscow State University \\
Moscow 119899, Russia}

\maketitle

\begin{abstract}
A brief review of recent results on
asymptotic expansions of Feynman  integrals
on the mass shell in momenta and masses
and their application to 2-loop calculations is
presented.
\end{abstract}

\section*{Introduction}

Explicit formulae for asymptotic expansions of Feynman diagrams
in various limits of momenta and masses,
with the large off-shell momenta,
have been derived in the simplest form
in \cite{vsmi:1,vsmi:2,vsmi:3} --- see a review, from the point
of view of 1995, in ref.~\cite{vsmi:4}.
(See also informal arguments in \cite{vsmi:5}.)

In this note, we present a brief review of recent results
obtained for two typical limits with the large momenta on
the mass shell: the limit of the large momenta on the mass shell
with the large mass (in Section~2) and the Sudakov limit, with the
large momenta on the massless mass shell (in Section~3).
In particular, the presented explicit formulae provide coefficients
at all powers and logarithms for two-loop diagrams in the Sudakov
limit.

\section*{Large momentum expansion on the mass shell}

Let us consider asymptotic expansion of a Feynman diagram
$F_{\Gm} (\uQ , \uq, \uM , \um)$ corresponding to a graph $\Gm$
in the limit when
the momenta $\uQ \equiv \{ Q_1, \ldots Q_i, \dots \}$ and
the masses $\uM \equiv \{ M_1, \ldots M_i, \dots \}$ are larger than
$\uq \equiv \{ q_1, \ldots q_i, \dots \}$ and
$\um \equiv \{ m_1, \ldots m_i, \dots \}$.
We suppose that the external momenta are non-exceptional and that
$\left(\sum_{i\in I} Q_i\right)^2 \neq M_j^2$, for any subset of indices $I$.
Moreover, let the large external momenta be on the mass shell, $Q_j^2=M_j^2$.

To obtain explicit formulae of the corresponding asymptotic expansion,
a method \cite{vsmi:6,vsmi:7}
based on constructing a remainder of the expansion, written through
an appropriate $R$-operation and using then diagrammatic Zimmermann identities
have been applied in refs.~\cite{vsmi:8,vsmi:9}.
The resulting explicit formula looks like
\be
F_{\Gm} (\uQ, \uM, \uq, \um;\ep)
\; \stackrel{\mbox{\footnotesize$M_j \to \infty$}}{\mbox{\Large$\sim$}} \;
\sum_{\gamma}  {\cal M}_{\gm}
F_{\Gm} (\uQ, \uM, \uq, \um;\ep) \, .
\label{E:vsmi:1}
\ee

The operator ${\cal M}_{\gm}$ happens
to be a product $\prod_i {\cal M}_{\gm_i}$ of
operators of Taylor expansion in certain momenta and masses.
For connectivity components $\gm_i$ other than $\gm_0$
(this is the component with the large external momenta), the
corresponding operator performs Taylor expansion of the Feynman integrand
$F_{\gm_i}$ in its small masses and external momenta.
The component $\gm_0$ can be naturally
represented as a union of its 1PI components and cut lines
(after a cut line is removed the subgraph becomes disconnected;
here they are of course lines with the large masses). By definition
${\cal M}_{\gm_0}$ is again factorized and the Taylor expansion of
the 1PI components of $\gm_0$ is performed as in the case of c-components
$\gm_i, \; i=1,2,\ldots$.

It suffices now to describe the action of the
operator ${\cal M}$ on the cut lines. Let $l$ be such a line,
with a large mass $M_j$, and let its momentum be $P_l+k_l$ where
$P_l$ is a linear combination of the large external momenta and
$k_l$ is a linear combination of the loop momenta and small external momenta.
If $P_l=Q_i$ then the operator $\cal M$ for this component of $\gm$ is
$\left. {\cal T}_{\ka} \frac{1}{\ka k_l^2+2Q_i k_l} \right|_{\ka=1} \, .$
(We use ${\cal T}_{\ldots}$ to denote
the operator of Taylor expansion at zero values
of the corresponding variables.)
In all other cases, e.g. when $P_l =0$, or it is a sum of two or more
external momenta, the Taylor operator ${\cal M}$ reduces
to ordinary Taylor expansion in small (with  respect to this line
considered as a subgraph) external momenta, i.e.
$\left.{\cal T}_{\ka} \frac{1}{ (\ka k_l+P_l)^2-M_i^2} \right|_{\ka=1}\,.$
Note that in all cases apart from the cut lines with $P_l^2 =M^2_j$
the action of the corresponding operator $\cal M$ is graphically
described (as for the off-shell limit) by contraction of the
corresponding subgraph to a point and
insertion of the resulting polynomial into the reduced vertex of the
reduced graph.

The formula (\ref{E:vsmi:1}) was illustrated in \cite{vsmi:8} through
a one-loop example and applied in \cite{vsmi:9} to calculation
of the master two-loop diagram shown in Fig.~1a (where thick (thin)
lines correspond to the mass $M$ ($m$) and the external momentum
is at $Q^2=M^2$). There are four subgraphs, Fig.~1(a--d),
that contribute to the general
formula.
\setlength {\unitlength}{1mm}
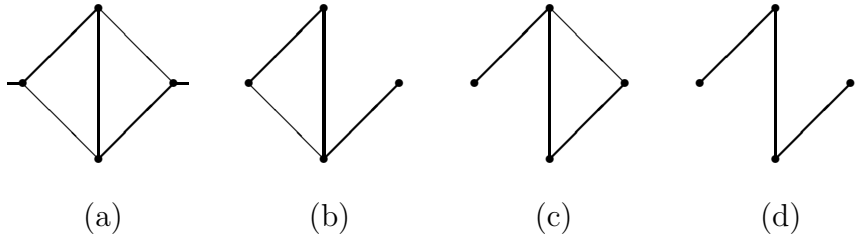
\begin{figure}[htb]
\begin{picture}(130,35)(-5,-5)
\put (10,20) {\circle*{1}}
\put (30,20) {\circle*{1}}
\put (20,10) {\circle*{1}}
\put (20,30) {\circle*{1}}
\thicklines
\put (10,20) {\line(1,1){10}}
\put (20,10) {\line(0,1){20}}
\put (20,10) {\line(1,1){10}}
\put (8,20) {\line(1,0){2}}
\put (30,20) {\line(1,0){2}}
\thinlines
\put (10,20) {\line(1,-1){10}}
\put (20,30) {\line(1,-1){10}}
\put (18,1) {(a)}
\put (40,20) {\circle*{1}}
\put (60,20) {\circle*{1}}
\put (50,10) {\circle*{1}}
\put (50,30) {\circle*{1}}
\thicklines
\put (40,20) {\line(1,1){10}}
\put (50,10) {\line(0,1){20}}
\put (50,10) {\line(1,1){10}}
\thinlines
\put (40,20) {\line(1,-1){10}}
\put (48,1) {(b)}
\put (70,20) {\circle*{1}}
\put (90,20) {\circle*{1}}
\put (80,10) {\circle*{1}}
\put (80,30) {\circle*{1}}
\thicklines
\put (70,20) {\line(1,1){10}}
\put (80,10) {\line(0,1){20}}
\put (80,10) {\line(1,1){10}}
\thinlines
\put (80,30) {\line(1,-1){10}}
\put (78,1) {(c)}
\put (100,20) {\circle*{1}}
\put (120,20) {\circle*{1}}
\put (110,10) {\circle*{1}}
\put (110,30) {\circle*{1}}
\thicklines
\put (100,20) {\line(1,1){10}}
\put (110,10) {\line(0,1){20}}
\put (110,10) {\line(1,1){10}}
\thinlines
\put (108,1) {(d)}
\end{picture}
\caption{(a) Two-loop master self-energy diagram.
(a--d) The subgraphs contributing to its large mass
expansions (1) at $q^2=M^2$.}
\end{figure}
 For example, the action of the operator ${\cal M}_{\gm}$
for $\gm=$Fig.~1d on the given Feynman integral
\be
\int\!\! \int \frac{\dd^d k \dd^d l}{(k^2-m^2)(l^2-m^2)
(k^2+2Qk)
\left[(k+l)^2 + 2Q(k+l)\right]
(l^2+2Ql)} .
\label{E:vsmi:2}
\ee
reduces to Taylor expansion of the last three factors in the
integrand, resp., in $k^2$, $(k+l)^2$, and $l^2$.
Each term of the corresponding asymptotic expansion happens to
be analytically computable --- see details and results in \cite{vsmi:9}.

Note that similar results for a simpler propagator-type 2-loop
diagram in the same limit were obtained and applied in
ref.~\cite{vsmi:10}. Further applications of the presented explicit
asymptotic expansions can be found in refs.~\cite{vsmi:11}.

\section*{Asymptotic expansion in the Sudakov limit}

One of possible versions of  the Sudakov limit is formulated
as the behaviour of a three-point Feynman diagram $F_{\Gm}(p_1,p_2,m)$
depending on two momenta, $p_1$ and $p_2$, on the massless mass shell,
$p_i^2=0$, with $q^2\equiv -Q^2=(p_1-p_2)^2 \to -\infty$.
We suppose, for simplicity, that there is one small non-zero mass, $m$.

To derive the simplest explicit formula for this limit,
the previous strategy of constructing
a remainder determined by an appropriate pre-subtraction operator and
then applying Zimmermann identities can be used.
Eventually, it takes an explicit form \cite{vsmi:8}
similar to (\ref{E:vsmi:1}), with another operator ${\cal M}_{\gm}$:
\be
 F_{\Gm}(p_1,p_2,m,\ep)
\; \stackrel{\mbox{\footnotesize$q^2 \to -\infty$}}{\mbox{\Large$\sim$}} \;
\sum_{\gamma}  {\cal M}_{\gm}
 F_{\Gm}(p_1,p_2,m,\ep) \, .
\label{E:vsmi:3}
\ee
Here the sum runs over subgraphs $\gm$ of $\Gm$
for which at least one of the following conditions holds:

({\em i}) In $\gm$ there is a path between the end-points 1 and 3.
(The end-points of the diagram are numerated according to
the following order: $p_1, p_2, q=p_1-p_2$.)
The graph $\hat{\gm}$ obtained from $\gamma$ by identifying
these end-points is 1PI.

({\em ii}) Similar condition with $1 \leftrightarrow 2$.

The pre-subtraction operator ${\cal M}_{\gm}$ is now
defined as a product $\prod_j {\cal M}_{\gm_j}$ of
operators of Taylor expansion acting on 1PI components and cut lines
of the subgraph $\gm$.
Suppose that the above condition ({\em i}) holds and ({\em ii})
does not hold.
Let $\gm_j$ be a 1PI component of $\gm$ and let $p_1+k$ be one
of its external momenta,
where $k$ is a linear combination of the loop momenta.
(We imply that the loop momenta are chosen
in such a way that $p_1$ flows through all $\gm_j$ and the
corresponding cut lines).
Let now $\uq_j$ be other independent external momenta of $\gm_j$.
Then the operator $\cal M$ for this component is defined as
${\cal T}_{k-((p_1 k)/(p_1p_2)) p_2,\uq_j, \um_j } \,$,
where $\um_j$ are the masses of $\gm_j$.
In other words, it is the operator of Taylor expansion in $\uq_j$ and
$\um_j$
at the origin and in $k$ at the point
$\tilde{k} =\frac{(p_1 k)}{(p_1p_2)} p_2$ (which depends on $k$ itself).

 For the cut lines the same prescription is adopted.
If $p_1 +k$ is the momentum of the cut line, then the
corresponding operator acts as
$\left. {\cal T}_{\ka} \frac{1}{\ka (k_l^2-m_l^2) +2 p_1 k }
\right|_{\ka=1} \, .$
If both ({\em i}) and ({\em ii}) hold the corresponding
operator performs Taylor expansion in the mass and the external momenta
of subgraphs (apart from $p_1$ and $p_2$).

The formula (\ref{E:vsmi:3}) was illustrated in \cite{vsmi:8} through
a one-loop example and applied in \cite{vsmi:12} to calculation
of the diagram shown in Fig.~2a in two cases: (a)
$m_1=\ldots =m_5=0, \;m_6=m$, and (b) $m_1=\ldots =m_4=0, \;m_5=m_6=m$.

For example, in the latter case of the Feynman integral
\bea
 F_i(p_1,p_2,m,\ep)
= \int  \int \frac{\dd^dk \dd^dl}{(k^2-2 p_1 k) (k^2-2 p_2 k) (k^2-m^2)}
\nn \\ \times
 \frac{1}{(l^2-2 p_1 l) (l^2-2 p_2 l) ((k-l)^2-m^2)} \, ,
\label{E:vsmi:4}
\eea
the subgraphs that give non-zero contributions
to the general formula (\ref{E:vsmi:3}) are shown in Fig.~2a-f.
The most complicated contribution comes from
Fig.~2e and~f, when the corresponding subtraction operator
expand propagators $1/((k^2-2 p_1 k) (l^2-2 p_1 l))$ (respectively,
$1/((k^2-2 p_2 k) (l^2-2 p_2 l))$) in Taylor series in $k^2$ and $l^2$.
In ref.~\cite{vsmi:12}, all the terms at the leading power were
analytically calculated, with the aid of Mellin--Barnes and
$\al$-parametric representations. Using the method of integration
by parts within dimensional regularization \cite{vsmi:13} one can arrive
at recurrence relations that provide all the relevant integrals
contributing to an arbitrary power of the expansion.
For example, here are results for the first four powers in
the expansion of this diagram:
\bea
-\frac{1}{\pi^4} (Q^2)^2 F_{\Gm} (Q^2,m^2)
\; \stackrel{\mbox{\footnotesize$Q^2 \to \infty$}}{\mbox{\Large$\sim$}} \;
\frac{1}{24} \ln^4 t + \frac{\pi^2}{3} \ln^2 t
- 6 \zeta(3) \ln t + \frac{31}{2} \zeta(4)
\nn \\
- \frac{1}{t} \left(
\frac{1}{6} \ln^3 t - \frac{3}{2} \ln^2 t
+ 6 \ln t  + \frac{2\pi^2}{3}\ln t
+18 - \frac{\pi^2}{3} -   6\zeta(3) \right)
\nn \\
- \frac{1}{t^2} \left(
\frac{1}{12} \ln^3 t - \frac{35}{8} \ln^2 t
+ \frac{29}{4} \ln t  + \frac{\pi^2}{3}\ln t
+ \frac{153}{8} - \frac{19\pi^2}{12} -   3\zeta(3) \right)
\nn \\
- \frac{1}{t^3} \left(
\frac{1}{18} \ln^3 t - \frac{55}{6} \ln^2 t
+ \frac{503}{36} \ln t  + \frac{2\pi^2}{9}\ln t
+ \frac{1061}{36} - \frac{355\pi^2}{108} -  2\zeta(3) \right) ,
\label{E:vsmi:5}
\eea
where $t=Q^2/m^2$.

\newpage

\setlength {\unitlength}{1pt}
\begin{figure}[t]
\begin{picture}(360,160)(-35,25)

\Line(10,100)(60,100)
\Line(10,150)(60,150)
\Line(10,100)(10,150)
\Line(60,150)(60,100)
\Line(60,100)(35,70)
\Line(35,70)(10,100)
\ArrowLine(35,50)(35,70)
\ArrowLine(10,150)(10,170)
\ArrowLine(60,170)(60,150)

\Text(35,45)[]{$p_1-p_2$}

\Text(10,175)[]{$p_1$}
\Text(60,175)[]{$p_2$}
\Text(35,25)[]{$(a)$}
\Text(  0,118)[]{3}
\Text( 25, 70)[]{1}
\Text( 73,118)[]{4}
\Text( 50, 70)[]{2}
\Text( 35,140)[]{5}
\Text( 35, 90)[]{6}
\Vertex(10,100){1.5}
\Vertex(60,100){1.5}
\Vertex(10,150){1.5}
\Vertex(60,150){1.5}
\Vertex(35,70){1.5}

\Line(120,150)(170,150)
\Line(120,100)(120,150)
\Line(170,150)(170,100)
\Line(145,70)(120,100)
\Line(170,100)(145,70)

\Vertex(120,100){1.5}
\Vertex(170,150){1.5}
\Vertex(145,70){1.5}
\Vertex(120,150){1.5}
\Vertex(170,100){1.5}

\Text(145,25)[]{$(b)$}


\Line(230,100)(230,150)
\Line(255,70)(230,100)
\Line(280,100)(255,70)

\Line(230,100)(280,100)

\Vertex(230,100){1.5}
\Vertex(230,150){1.5}
\Vertex(280,100){1.5}
\Vertex(255,70){1.5}

\Text(255,25)[]{$(c)$}


\Line(360,150)(360,100)
\Line(335,70)(310,100)
\Line(360,100)(335,70)

\Line(310,100)(360,100)

\Vertex(310,100){1.5}
\Vertex(360,100){1.5}
\Vertex(360,150){1.5}
\Vertex(335,70){1.5}

\Text(335,25)[]{$(d)$}

\end{picture}

\begin{picture}(360,140)(-35,12)


\Line(135,100)(135,150)
\Line(160,70)(135,100)
\Text(160,25)[]{$(e)$}
\Vertex(135,100){1.5}
\Vertex(135,150){1.5}
\Vertex(160,70){1.5}


\Line(265,150)(265,100)
\Line(265,100)(240,70)
\Text(240,25)[]{$(f)$}
\Vertex(265,100){1.5}
\Vertex(265,150){1.5}
\Vertex(240,70){1.5}

\vspace{-10pt}

\end{picture}
\caption {(a) A typical two-loop vertex diagram. (a)--(f)
Subgraphs contributing to the asymptotic expansion
of the diagram (a) in the Sudakov limit.}
\end{figure}
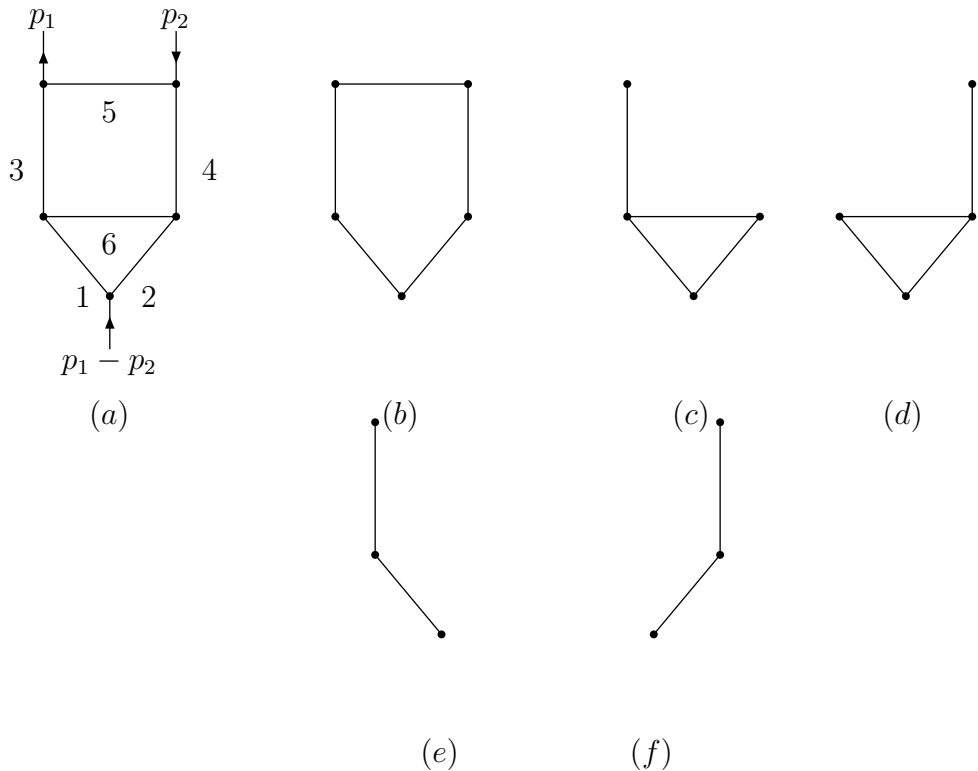

Note that a typical feature of the off-shell explicit
formulae is an interplay
between ultraviolet and infrared divergences which appear
in individual terms of the expansion but are mutually canceled,
provided the initial Feynman integral is finite. It turns out that,
for the Sudakov limit, one meets a more general interplay
between ultraviolet, collinear and infrared divergences which,
at first sight, seem to be of very different nature.
In particular, in the case of the Feynman integral (\ref{E:vsmi:4}),
which does not have divergences from the very beginning,
individual terms of the corresponding asymptotic
expansion involve all the three kinds of divergences resulting in
poles up to $1/\eps^4$, with $\eps = (4-d)/2$ parameter of dimensional
regularization \cite{vsmi:14}.
Namely, the contribution of Fig.~2a (as a subgraph)
involves infrared and collinear divergences,
the contribution of Fig.~2(e\&f) involves ultraviolet and collinear
divergences, while other three contributions, Fig.~2b and
Fig.~2(c\&d) possess all kinds of the divergences.
However these poles are successfully canceled in the sum
and one obtains expansion (\ref{E:vsmi:5}) in the limit
$\eps \to 0$.

Note that similar explicit formulae can be derived and applied
for other versions of the Sudakov limit, e.g.
$p_i^2=m^2$, with $q^2\equiv (p_1-p_2)^2 \to -\infty$.
It is natural to expect that
the presented possibility to obtain all powers and logarithms
in two loops can be applied to
extend well-known results on asymptotic behaviour in the Sudakov
limit in QED and QCD \cite{vsmi:15}.

Extensions of the presented methods and results to other
typically Minkowskian limits and their applications
will be reported elsewhere.

\vspace{1mm}

My participation in the Balholm Workshop has been supported by
the Organizing Committee of the workshop
and by the Russia Ministry for Science and Technologies.
The research has been supported by the Russian Foundation for Basic
Research, project 96--01--00654, and by INTAS, project 93--0744.

\vspace{1mm}

{\em Acknowledgments.}
I am grateful to P. Osland for kind hospitality during the workshop
and to K.G.~Chetyrkin, A.~Czarnecki, A.I.~Davydychev,
K.~Melnikov and J.B.~Tausk for helpful discussions.


\begin{references}
\bibitem{vsmi:1}
Gorishny, S.G., {\it Nucl. Phys.} {\bf B319}, 633 (1989).

\bibitem{vsmi:2}
Chetyrkin, K.G., {\it Teor. Mat. Fiz.} {\bf 75}, 26; {\bf 76},  207 (1988);
preprint MPI-PAE/PTh 13/91 (Munich, 1991).

\bibitem{vsmi:3}
Smirnov, V.A., {\it Commun. Math. Phys.} {\bf 134},  109 (1990);
{\it Renormalization and asymptotic expansions},
Basel: Birkh\"{a}user, 1991.

\bibitem{vsmi:4}
Smirnov, V.A., {\it Mod. Phys. Lett.} {\bf A 10}, 1485 (1995).

\bibitem{vsmi:5}
Larin, S.A., van Ritbergen, T., and Vermaseren, J.A.M.,
{\it Nucl.~Phys.} {\bf B438}, 278 (1995).

\bibitem{vsmi:6}
Zimmermann, W., {\it Commun. Math. Phys.} {\bf 15}, 208 (1969);
{\it Ann. Phys.} {\bf 77}, 570 (1973).

\bibitem{vsmi:7}
Anikin, S.A., and Zavialov, O.I., {\it Teor. Mat. Fiz.} {\bf 27}, 425 (1976);
{\it Ann. Phys.} {\bf 116}, 135 (1978);
Zavialov, O.I., {\it Renormalized Quantum Field Theory}, Kluwer
Academic  Publishers, 1990.

\bibitem{vsmi:8}
Smirnov, V.A., {\it Phys. Lett.} {\bf B394}, 205 (1997).

\bibitem{vsmi:9}
Czarnecki, A., and Smirnov, V.A., {\it Phys. Lett.} {\bf B394}, 211 (1997).

\bibitem{vsmi:10}
Avdeev, L.V., and Kalmykov, M.Yu., hep-ph/9701308, to appear in
{\it Nucl. Phys~B}.

\bibitem{vsmi:11}
Czarnecki, A., and Melnikov, K.,
Phys. Rev. Lett. {\bf 78}, 3630 (1997); hep-ph/9706227;
Czarnecki, A., Melnikov, K., and Uraltsev, N.,
hep-ph/9706311.

\bibitem{vsmi:12}
Smirnov, V.A., hep-ph/9703357, to appear in {\it Phys. Lett. B.}

\bibitem{vsmi:13}
Chetyrkin, K.G., and Tkachov, F.V., {\it Nucl. Phys.} {\bf B192}, 159 (1981).

\bibitem{vsmi:14}
't Hooft, G., and Veltman, M., {\it Nucl.~Phys.} {\bf B44}, 189 (1972);
Bollini, C.G., and Giambiagi, J.J., {\it Nuovo Cim.} {\bf 12B}, 20 (1972).

\bibitem{vsmi:15}
Sudakov, V.V., {\it Zh. Eksp. Teor. Fiz.} {\bf 30}, 87 (1956);
Cornwall, J.M., and Tiktopoulos, G., {\it Phys. Rev.}
{\bf D13}, 3370 (1976);
Belokurov, V.V., and Ussyukina, N.I., {\it Teor. Mat. Fiz.}
{\bf 47}, 157 (1979); {\bf 44}, 147 (1980);
Collins, J.C., in {\it Perturbative QCD}, ed. Mueller, A.H., 1989, p.~573;
Korchemsky, G.P., {\it Phys. Lett.} {\bf B217}, 330 (1989); {\bf B220},
629 (1989).
\end{references}
\end{document}